\documentclass[onecolumn,letterpaper,amsmath,amssymb,floatfix,aps,superscriptaddress]{revtex4}

\def\ul#1#2{\textstyle{\frac{#1}{#2}}}

\newcommand {\bnabla} {\mbox{\boldmath$\nabla$}}

\usepackage{graphicx}
\usepackage{dcolumn}  
\usepackage{bm}  
\usepackage{epsfig}

\begin{document}
\setlength\arraycolsep{2pt}

\title{Ordering of anisotropic polarizable polymer chains on the full many-body level}

\author{David S. Dean}
\affiliation{Universit\'e de  Bordeaux and CNRS, Laboratoire Ondes et Mati\`ere d'Aquitaine (LOMA), UMR 5798, F-33400 Talence, France}
\affiliation{ Laboratoire de
Physique Th\'eorique (IRSAMC),Universit\'e de Toulouse, UPS and CNRS,  F-31062 Toulouse, France}

\author{Rudolf Podgornik}
\affiliation{Department of Theoretical Physics, J. Stefan
Institute, SI-1000 Ljubljana, Slovenia} 
\affiliation{Department of Physics, Faculty
of Mathematics and Physics, University of Ljubljana, SI-1000
Ljubljana, Slovenia}
\affiliation{ Laboratoire de
Physique Th\'eorique (IRSAMC),Universit\'e de Toulouse, UPS and CNRS,  F-31062 Toulouse, France}

\begin{abstract}
We study the effect of dielectric anisotropy of polymers on their equilibrium ordering within mean-field
theory but with a formalism that takes into account the full n-body nature of van der Waals forces.
Dielectric anisotropy within polymers is to be expected as the electronic properties of the polymer will
typically be different along the polymer than across its cross section. It is therefore physically intuitive that larger charge fluctuations can be induced along the chain than perpendicular to it. We show that this dielectric  anisotropy leads to n-body interactions which can induce an isotropic--nematic transition. The two body and three body components of the full van der Waals interaction are extracted and it is shown how the 
two body term behaves like the phenomenological self-aligning-pairwise nematic interaction.
At the three body interaction level we see that the nematic phase 
that is energetically favorable is discotic, however on the full n-body interaction level 
we find that the normal axial nematic phase is always the stable ordered phase.  The n-body nature of
our approach also shows that the key parameter driving the nematic-isotropic transition is the bare
persistence length of the polymer chain. 
\end{abstract}
\maketitle


\section{Introduction}

Semiflexible polymers, just like rod-like anisotropic nematogens, form orientationally ordered liquid crystalline mesophases, spanning the concentration regime of chiral nematics \cite{Rill} and all the way to line hexatics \cite{Podgornik-hexatic}. Among the most important examples of semi flexible polymer ordering is the double stranded DNA molecule \cite{Livolant, Podgornik-rev1} which makes a plethora of liquid crystalline phases in aqueous solutions. These ordered phases  of DNA aqueous dispersions are pertinent to the construction of DNA nanoassemblies \cite{Evdokimov} and have recently found a promising new application in the form of virus-like particles \cite{Steinmetz}.

A rigorous theory of nematic ordering of long rigid nematogens  in dilute solutions was first proposed in a seminal work by Onsager \cite{Onsager}. His  approach was later adapted to semiflexible polymers by Semenov and Khokhlov \cite{Semenov} and within a different formal background by Ronca and Yoon \cite{Yoon}. A field-theoretical reformulation of the nematic-isotropic transition for semi flexible polymers was then pursued by Gupta and Edwards \cite{Edwards} as well as by Tkachenko and Rabin \cite{Tkachenko}. These approaches are all based on the assumption of pairwise additivity of the interaction potential, assumed to be of a general nematic form stemming from the Onsager steric interaction. For short range interactions this is of course a reasonable assumption. However, it is well known that 
electrostatic as well as van der Waals (vdW) interactions usually engender long range non-pairwise additive interactions that in general do not conform to this approximation \cite{revmod}.

Furthermore, polymers in general show an anisotropic dielectric response, so that the dielectric functions along the axis and perpendicular to the axis differ. This is nicely seen {\sl e.g.} in the case of DNA \cite{MacNaughton}, where amplitude of the dielectric function in the axial and the radial directions are {\sl very} different because of the strong optical anisotropy of DNA \cite{Ching}. This dielectric anisotropy has important consequences as vdW interactions between anisotropic media lead in general to torques induced by electromagnetic field fluctuations as first realized by Weiss and Parsegian \cite{Weiss,measuretorque}. The details of this effect are complicated but it persists not only in the non-retarded but also in the retarded limit \cite{Barash,Siber1,Siber2}. 

Existence of anisotropic vdW interactions now poses a question about their possible role in ordering transitions. While the fundamental importance of the vdW force for the gas-liquid transition is well recognized we now investigate the possible role of anisotropic vdW interactions for orienational ordering transitions. As it is not clear off-hand whether the effect is significant, it is of utmost importance that the many-body pairwise non-additive nature of vdW interactions be fully recognized in the theoretical framework. We thus formulate the theory of orientational ordering of semi flexible polymers due to anisotropic vdW interactions in the way that allows for a complete and exact resummation of vdW interaction energy to all levels, including pairs, triplets, quadruplets etc. Whatever the conclusions of this theory, they can certainly not be undermined by the improper treatment of the vdW effect. 

In order to assess the effect of anisotropic vdW interactions we analyze a rather simple case of semiflexible polymer ordering 
driven exclusively by these interactions. We thus study the effects beyond the usual Onsager steric anisotropy {\sl ansatz}. However, the most 
fundamental difference with the previous work is that we calculate {\sl exactly} the contribution of vdW interactions
to all orders within the mean-field approximation. Usually vdW interactions are treated on a pairwise level, or possibly three body Axilrod-Teller level \cite{Ninham}, but never more. We propose an analytical approach that allows us to sum the van der Waals interactions to all orders and solve the corresponding mean-field theory exactly. 
It should be noted that  the long range nature of van der Waals interactions implies that our mean-field treatment  should be reasonably accurate, especially in dense systems.  Pursuing this approach we show  indeed that the anisotropic vdW interactions can engender an orientational ordering transition in a polymer solution of stiff chains.

\section{Electrostatic field energy}

We start with a polymer described as a continuous chain within the arclength parametrization as ${\bf x}(s)$ so that the tangent vector is defined as $${\bf t}(s) = \dot{\bf x}(s).$$The chain is assumed to be composed of an anisotropic macroscopic dielectric material that has a dielectric response function $\varepsilon_{\parallel}$ and $\varepsilon_{\perp}$ parallel and perpendicular to the chain, respectively. The electrostatic energy of the polymer-solvent system is thus described as
\begin{equation}
{\cal E}_{ES} = {\ul12}\varepsilon_w \int_{V_w} \left( \bnabla \phi({\bf x})\right)^2 d{\bf x} + {\ul12}\varepsilon_{\parallel} \int_{V_p} t_i t_j  \nabla_i \phi({\bf x}) \nabla_j \phi({\bf x}) d{\bf x} + {\ul12}\varepsilon_{\perp} \int_{V_p} \left( \delta_{ij} - t_i t_j \right) \nabla_i \phi({\bf x}) \nabla_j \phi({\bf x}) d{\bf x}.
\end{equation}
Assuming now that the polymer is much thinner, of radius $a$,  then its length we get the final expression of the electrostatic energy as
\begin{equation}
{\cal E}_{ES} = {\ul12}\varepsilon_w \int_{V} \left( \bnabla \phi({\bf x})\right)^2 d{\bf x} + {\ul{\pi a^2}{2}} \int_{{\bf x}(s)} ds \big( \left( \varepsilon_{\perp} -  \varepsilon_w\right) \bnabla \phi({\bf x}(s))^2 + \left( \varepsilon_{\parallel} -  \varepsilon_{\perp}\right)
t_i t_j  \nabla_i \phi({\bf x}(s)) \nabla_j \phi({\bf x}(s)) \big).
\end{equation}
Here the first integral is now over the whole space and the second one is a line integral along the polymer chain. We introduce the scalar polymer density field $\rho({\bf x})$ and the tensor polymer nematic field $\sigma_{ij}({\bf x})$ defined as
\begin{equation}
\rho({\bf x}) =  \int_0^L ds ~\delta({\bf x} - {\bf x}(s)), 
\end{equation}
and
\begin{equation}
\sigma_{ij}({\bf x}) =  \int_0^L ds ~\delta({\bf x} - {\bf x}(s)) t_i({\bf x}(s)) t_j({\bf x}(s)), 
\end{equation}
where $L$ is the total length of the polymer. We can now write the electrostatic energy as
\begin{equation}
{\cal E}_{ES}[\phi({\bf x})] = {\ul12} \int_{V} \tilde\varepsilon_{ij}({\bf x}) \nabla_i \phi({\bf x}) \nabla_j \phi({\bf x}),
\end{equation}
where the composite dielectric function  is now given by
\begin{equation}
\tilde\varepsilon_{ij}({\bf x}) = \left( \varepsilon_w + \pi a^2 ( \varepsilon_{\perp} -  \varepsilon_w)~ \rho({\bf x}) \right) \delta_{ij} +  \pi a^2 (\varepsilon_{\parallel} -  \varepsilon_{\perp}) ~\sigma_{ij}({\bf x}).
\label{coupling}
\end{equation}
In thermodynamic equilibrium electric field fluctuations give rise to the partition function
\begin{equation}
{\cal Z}_{vdW} = \int {\cal D}[\phi({\bf x})]~\exp{(- \beta {\cal E}_{ES}[\phi({\bf x})])}.
\end{equation}
Since the integral over the electrostatic field, in the limit where the polymer fields are both independent and constant {\sl i.e.} $\rho({\bf x}) = \rho $ and $\sigma_{ij}({\bf x}) = \sigma_{ij}$ , is of a general Gaussian type it can be evaluated explicitly giving for the corresponding free energy 
\begin{equation}
\beta {\cal F}_{vdW} = - \log{{\cal Z}_{vdW}} = {\ul12} V \int \frac{d^3 {\bf k}}{(2\pi)^3} \log{\left( \tilde\varepsilon_{ij} k_i k_j \right)}.
\end{equation}
This is the form of the free energy corresponding to thermal fluctuations of the electrostatic field that we will use later. It is equivalent to the zero Matsubara frequency or static van der Waals interactions as is well known \cite{Ninham}. In the full Lifshitz theory \cite{vdWbooks} with assumed complete magnetic homogeneity of the space, the above formula is generalized to 
\begin{equation}
\beta {\cal F}_{vdW} = - \log{{\cal Z}_{vdW}} =  V {\sum_{n = 0}^{\infty}}' \int \frac{d^3 {\bf k}}{(2\pi)^3} \log{\left( \tilde\varepsilon_{ij}(\imath \xi_n) k_i k_j \right)},
\label{fullmats}
\end{equation}
where the $n$ summation (the prime indicates that $n = 0$ term has a weight of $1/2$) is over the imaginary Matsubara frequencies $\xi_n = 2\pi n k_BT/\hbar$, where $k_B$ is the Boltzman constant, $T$ is the absolute temperature, and $\hbar$ is the Planck constant divided by $2\pi$. 

\section{Polymer energy}

The polymer chain itself is not featureless and will be modeled within the framework of the semiflexible Kratky-Porod model where the polymer conformational energy is given by
\begin{equation}
{\cal E}_{KP} = {\ul12} K \int_0^L ds~ {\dot{\bf t}}(s)^2 \qquad {\rm with ~the~constraint} \qquad {\bf t}(s)^2 = 1,
\end{equation}
where the last constraint stems from the fixed length of the monomers. Instead of dealing with this constraint locally, which is a difficult task, we simplify the model as suggested by Edwards and Gupta \cite{Edwards}, by taking it into account globally in the form 
\begin{equation}
\left< {\bf t}(s)^2 \right> = 1.
\label{cons}
\end{equation}
This then leads to the effective partition function of the polymer chain in the form
\begin{equation}
{\cal E}_{KP} = {\ul12} K \int_0^L {\dot{\bf t}}(s)^2 ds + {\ul12} \lambda \int_0^L \left( {\bf t}(s)^2 - 1\right) ds,
\end{equation}
where $\lambda$ is a Lagrange multiplier to be evaluated self-consistenly from the global constraint of Eq. (\ref{cons}). The polymer partition function is then given by the form
\begin{equation}
{\cal Z}_{KP} = \int {\cal D}[{\bf x}(s)]~\exp{(- \beta {\cal E}_{KP}[{\bf x}(s)])} \qquad {\rm with} \qquad \int {\cal D}[{\bf x}(s)]~{\dot {\bf x}}(s)^2 \exp{(- \beta {\cal E}_{KP}[{\bf x}(s)])} = {\cal Z}_{KP}.
\end{equation}

\section{Constraint "energy"}

In order to evaluate the total free energy of the system we need to insert the proper constraints into the total energy of the fluctuating electrostatic field and polymer configurations. That the two are coupled is evident from Eq. (\ref{coupling}). In the final partition function one thus needs to insert the following delta functions that can be represented with the functional delta transforms
\begin{equation}
\int {\cal D}[{\rho}({\bf x})]~ \delta\left( \rho({\bf x}) - \int_0^L\!\!\!\!\!\!ds ~\delta({\bf x} - {\bf x}(s))\right) = \int {\cal D}[{\rho}({\bf x})] {\cal D}[{r}({\bf x})] \exp{\left( \imath \int_V \!\!\!d{\bf x}~ {r}({\bf x}) {\rho}({\bf x}) - \imath \int_0^L \!\!\!\!\!\!ds~ {r}({\bf x}(s)) \right)},
\end{equation}
where $ {\bf r}({\bf x})$ is the corresponding auxiliary field, and
\begin{eqnarray}
& & \int {\cal D}[\sigma({\bf x}) ] ~\delta\left( \sigma_{ij}({\bf x}) -\int_0^L \!\!\!\!\!\!ds ~\delta({\bf x} - {\bf x}(s)) t_i({\bf x}(s)) t_j({\bf x}(s))\right)  = \nonumber\\
& & \int {\cal D}[\sigma({\bf x}) ] {\cal D}[s({\bf x})] \exp{\left( \imath \int_V \!\!\!d{\bf x}~ s_{ij}({\bf x}) {\sigma}_{ij}({\bf x}) - \imath \int_0^L \!\!\!\!\!\!ds~ s_{ij}({\bf x}(s)) t_i({\bf x}(s)) t_j({\bf x}(s))\right)},
\end{eqnarray}
where the auxiliary field is now the tensor $s({\bf x})$ with components $s_{ij}({\bf x})$ and we adopt the Einstein summation convention. When using these constraints in the compact definition of the complete partition function we will also perform a Wick rotation in the complex plane for the two auxiliary fields as well as redefine both fields in such a way that they scale with the inverse temperature. Since they are arbitrary fields anyway this does not introduce any inconsistencies.

These two constraints can then be rewritten in a form of an effective "constraint partition function" of the form
\begin{equation}
{\cal Z}_{CO} = \int {\cal D}[{\rho}({\bf x})] {\cal D}[{r}({\bf x})] {\cal D}[\sigma({\bf x}) ] {\cal D}[s({\bf x})]  \exp{\left( - \beta {\cal E}_{CO}[{r}({\bf x}), {\rho}({\bf x}), s_{ij}({\bf x}), \sigma_{ij}({\bf x})])\right)},
\end{equation}
where
\begin{eqnarray}
{\cal E}_{CO}[{r}({\bf x}), {\rho}({\bf x}), s({\bf x}), \sigma({\bf x})] &=& - \int_V \!d{\bf x}~ {r}({\bf x}) {\rho}({\bf x}) -  {\ul12} \int_V \!d{\bf x}~ s_{ij}({\bf x}) {\sigma}_{ij}({\bf x})  + \nonumber\\
& & + \int_0^L\!\!\!\!\!\!ds~ {r}({\bf x}(s)) + {\ul12} \int_0^L \!\!\!\!\!\!ds~ s_{ij}({\bf x}(s)) t_i({\bf x}(s)) t_j({\bf x}(s)).
\end{eqnarray}

\section{Total partition function}

In view of the above the total partition function can be cast into the following form 
\begin{equation}
{\cal Z}_{T} = \int {\cal D}[\phi({\bf x})] {\cal D}[{\bf x}(s)] \int  {\cal D}[{\rho}({\bf x})] {\cal D}[{r}({\bf x})] {\cal D}[\sigma({\bf x}) ] {\cal D}[s({\bf x})] \exp{\left( - \beta {\cal E}_T\right)},
\end{equation}
where
\begin{eqnarray}
{\cal E}_T &=&  {\ul12} \int_{V} \tilde\varepsilon_{ij}({\bf x}) \nabla_i \phi({\bf x}) \nabla_j \phi({\bf x}) + \nonumber\\
& &  + {\ul12} K \int_0^L{\ddot{\bf x}}(s)^2 ds + {\ul12} \lambda \int_0^L \left( \dot{\bf x}(s)^2 - 1\right) ds + {\ul12} \int_0^L\!\!\!\!\!\!ds~ s_{ij}({\bf x}(s)) t_i({\bf x}(s)) t_j({\bf x}(s)) + \int_0^L \!\!\!\!\!\!ds~ {r}({\bf x}(s)) \nonumber\\
& & - \int_V \!d{\bf x}~ {r}({\bf x}) {\rho}({\bf x}) -  {\ul12} \int_V \!d{\bf x}~ s_{ij}({\bf x}) {\sigma}_{ij}({\bf x}).
\label{totaleber}
\end{eqnarray}
We now assume that both the polymer density field and the polymer orientational field are constant in space, which affectively means that we introduce a mean-field {\sl ansatz}. In that case we already evaluated the functional integral over $\phi({\bf x})$. We next tackle the polymer part being a functional integral over the polymer configurational field ${\bf x}(s)$.

In order to do that we first of all introduce the Rouse or polymer Fourier decomposition as
\begin{equation}
{\bf x}(s) = \int \frac{d\omega}{2\pi} {\bf x}(\omega) e^{-\imath \omega s}.
\end{equation}
The complete polymer part of the partition function can thus be written as
\begin{equation}
{\cal Z}_{{\bf x}(s)} = \exp{\left( {\ul12} \beta \lambda L - \beta r L\right)} ~\prod_{\omega} \int d{\bf x}(\omega) \exp{\left( - \beta \tilde{\cal E}_{PO}({\bf x}(\omega))\right)} 
\end{equation}
where $L$ is the total length of the polymer chain, {\sl i.e.} $L = N b$ and 
\begin{equation}
\tilde{\cal E}_{PO}({\bf x}(\omega)) =  {\ul12} a_{ij}(\omega) ~{x}_i(\omega) {x^*}_j(\omega),
\end{equation}
with
\begin{equation}
a_{ij}(\omega) =  \omega^2 \left( (K \omega^2 + \lambda) \delta_{ij}  + s_{ij}\right).
\end{equation}
The functional integral over the Rouse modes is now Gaussian and can be evaluated explicitly yielding up to irrelevant multiplicative constants
\begin{equation}
{\cal Z}_{{\bf x}(s)} = \frac{\exp{\left( {\ul12} \beta \lambda L - \beta r L\right)}}{\left( \det{a_{ij}(\omega)}\right)^{1/2}}.
\end{equation}
The intermediate partition function has thus been derived in the form
\begin{equation}
\tilde{\cal Z}_{T} = \int {\cal D}[{\rho}({\bf x})] {\cal D}[{r}({\bf x})] {\cal D}[\sigma({\bf x}) ] {\cal D}[s({\bf x})] \exp{\left( - \beta \tilde{\cal F}_T\right)},
\end{equation}
where the intermediate free energy entering the above definition can be written as
\begin{equation}
\beta \tilde{\cal F}_T = {\ul12} V \int \frac{d^3 {\bf k}}{(2\pi)^3} \log{\left( \tilde\varepsilon_{ij} k_i k_j \right)} + {\ul12} L   \int \frac{d\omega}{2\pi} \log{\det a_{ij}(\omega)} - \beta V~ {r}  {\rho}  - \beta V~ {\ul12} s_{ij}  {\sigma}_{ij} + \beta r L - {\ul12} \beta \lambda L .
\end{equation}
In order to derive this form we had to assume that all the relevant fields, {\sl i.e.} ${r}, {\rho}, s_{ij}$ and $ {\sigma}_{ij}$,  are constant in space.  Taking into account the connection between the monomer density $\rho$ and the length of the polymer chain, {\sl i.e.}
\begin{equation}
\frac{L}{V} = \frac{N b}{V} = n b \equiv \rho,
\end{equation}
we derive the intermediate free energy density as 
\begin{equation}
\frac{\beta \tilde{\cal F}_T({r}, {\rho}, s, {\sigma})}{V} = {\ul12} \int \frac{d^3 {\bf k}}{(2\pi)^3} \log{\left( \tilde\varepsilon_{ij}(\rho, \sigma_{ij}) k_i k_j \right)} + {\ul12} n b   \int \frac{d\omega}{2\pi} \log{\det a_{ij}(\omega, s_{ij})} - \beta \left( {r}  {\rho}  + {\ul12} s_{ij}  {\sigma}_{ij}\right)  + \beta \left(r - {\ul12}  \lambda \right) n b.
\end{equation}
with
\begin{equation}
\tilde\varepsilon_{ij}(\rho, \sigma) = \left( \varepsilon_w + \pi a^2 ( \varepsilon_{\perp} -  \varepsilon_w)~ \rho\right)  \delta_{ij} +  \pi a^2 (\varepsilon_{\parallel} -  \varepsilon_{\perp}) ~\sigma_{ij} \qquad {\rm and} \qquad a_{ij}(\omega, s) =  \omega^2 \left( (K \omega^2 + \lambda) \delta_{ij}  + s_{ij}\right).
\label{coupling1}
\end{equation}
As indicated, this still has to be evaluated in the thermodynamic equilibrium as a function of all four polymer fields. 

\section{Mean-field equations}

The second integral can be evaluated easily in an explicit form in terms of the eigenvalues $s_\mu$ 
of $s$ and $\sigma_\mu$ of $\sigma$, yielding for the free energy density
\begin{equation}
\frac{\beta \tilde{\cal F}_T({r}, {\rho}, s, {\sigma})}{V} = {\ul12} \int \frac{d^3 {\bf k}}{(2\pi)^3} \log{\left( \sum_{\mu}\tilde\varepsilon_{\mu}(\rho, \sigma_{ij}) k^2_{\mu}\right)} + {\ul14} n b \sum_{\mu} \sqrt{\frac{\lambda + s_{\mu}}{K}}  -  {\ul12} \beta \left( \sum_{\mu} s_{\mu}  {\sigma}_{\mu} +\lambda~nb\right) + \beta r (nb - \rho).
\end{equation}
The minimization is now carried out  with respect to all the fields, {\sl i.e.} $s, \sigma, \rho, r, \lambda$. 
Notice that the first term in the above is the term due to the interactions between the polymer chain
with itself, however the following terms are just the free energy for a polymer which is suitably constrained. The minimization with respect to the constraint parameters $s_\mu$, $r$ and $\lambda$
gives 
\begin{equation}
\frac{\partial (\beta \tilde{\cal F}_T)}{\partial \lambda} = {\ul14} \frac{nb}{\sqrt{K}} \sum_{\mu}\frac{1}{\sqrt{\lambda + s_{\mu}}} - \beta nb = 0 \qquad {\rm and} \qquad \frac{\partial (\beta \tilde{\cal F}_T)}{\partial r} = nb - \rho = 0,
\label{lamequ1}
\end{equation}
\begin{equation}
\frac{\partial (\beta \tilde{\cal F}_T)}{\partial s_{\mu}} = {\ul14} \frac{nb}{\sqrt{K}}\frac{1}{\sqrt{\lambda + s_{\mu}}} - \beta \sigma_{\mu} = 0 .
\label{lamequ2}
\end{equation}
These equations can now be used to eliminate the variables $s_\mu$ and $\rho$, the resulting free
energy being independent of $r$. The first equation in Eq. (\ref{lamequ1}) along with that 
of Eq. (\ref{lamequ2}) implies that $\sigma$ must obey the constraint
\begin{equation}
\sum_{\mu} \sigma_{\mu}  = {\rm Tr} \ \sigma = \rho = nb.
\end{equation}
The resulting effective free energy is thus a function solely of $\sigma$ and some simple algebra gives
\begin{equation}
{\beta \tilde{\cal F}_T(\sigma)\over V} =   {\beta \tilde{ \cal F}_{vdW}(\sigma) \over V} + {(n b)^2 k_BT \over 32~ K}
~{\rm Tr} ~\sigma^{-1}
\end{equation}

The zero Matsubara frequency van der Waals part of the free energy can be written up to terms independent of $\sigma$ as
\begin{equation}
{\beta \tilde{ \cal F}_{vdW} \over V} = {1\over 2} \int {d^3{\bf k}\over (2\pi)^3}[ \ln\left( \varepsilon_w + \pi a^2 ( \varepsilon_{\perp} -  \varepsilon_w)~ nb\right) +\ln( k^2 + \alpha \sigma_{ij} k_i k_j)]
\end{equation}
with 
\begin{equation}
\alpha = {\pi a^2 (\varepsilon_{\parallel} -  \varepsilon_{\perp})\over \varepsilon_w + \pi a^2 ( \varepsilon_{\perp} -  \varepsilon_w) nb},
\end{equation}
and we have used the relation $\rho=nb$.

It now remains to compute the zero Matsubara frequency van der Waals free energy which means we must compute an integral of
the form 
\begin{equation}
W(B) = \int d^3{\bf k}\  \ln( B_{ij}k_i k_j).
\end{equation}
To proceed we notice that 
\begin{equation}
 {\partial W(B)\over \partial B_{pq}} =  I_{pq} \int_0^\Lambda k^2 dk = I_{pq} {\Lambda^3\over 3}
 \end{equation}
 where
 \begin{equation}
 I_{pq} =\int d{\hat{\bf k}} {{\hat k}_p {\hat k}_q\over B_{ij} {\hat k}_i{\hat
k_j}}.
\end{equation}
Here  $\hat {\bf k}$ denotes the unit vector and $\Lambda$ is a ultraviolet cut-off corresponding to
length scales below which the electromagnetic field fluctuations are cut off.  We now note that 
\begin{equation}
J_{pq}(B) = \int d^3{\bf k} \exp(-{1\over 2} {\bf k}^2) {k_p k_q \over  B_{ij}{k}_i {k}_j} = \int_0^\infty k^2 dk \exp(-{1\over 2} k^2) I_{pq}(B_{ij}) = \sqrt{\pi\over 2} \ I_{pq}(B_{ij}).
\end{equation}
and use the identity
\begin{eqnarray}
J_{pq}(B)&=&  \int d^3{\bf k} \exp(-{1\over 2} {\bf k}^2) k_p k_q \int_0^\infty {dt\over 2}\exp(-{t\over 2}  B_{ij}{k}_i {k}_j) = {1\over 2} (2\pi)^{3\over 2}\int_0^\infty dt\  (I+ t B)_{pq}^{-1} \det( 1+ t B)^{-{1\over 2}},
\end{eqnarray}
where $I$ denotes the identity matrix.  Putting all of this together we obtain
\begin{equation}
I_{pq}(B) = 2\pi \int_0^\infty dt\ (1+t B)_{pq}^{-1} \det( I+ t B)^{-{1\over 2}}  = \frac{\partial W(B)}{ \partial B_{pq}}.
\end{equation}
Thus  we derive the  effective potential $W(B)$ in the form of
\begin{equation}
W(B = I + \alpha \sigma) = 4\pi \int_0^\infty {dt\over t} \ \left[ {1\over (1+t)^{3\over 2}}-\det(I+ t B)^{-{1\over 2}} \right] = 4\pi \int_0^\infty {dt\over t} \ \left[ {1\over (1+t)^{3\over 2}}-\det( I+t(I+\alpha \sigma))^{-{1\over 2}} \right] ,
\end{equation}

Using this result the zero Matsubara frequency van der Waals part of the  free energy can then be written in a compact form as 
\begin{equation}
{\beta \tilde{ \cal F}_{vdW}(\sigma) \over V}= {\Lambda^3\over 12 \pi^2}\left[ \ln\left(\varepsilon_w + \pi a^2 (\varepsilon_\perp -\varepsilon_w)nb\right) + \int_0^\infty {dt\over t}\left( {1\over (1+t)^{3\over 2}}
-\det[I+t(I+ \alpha \sigma)]^{-{1\over 2}}\right)\right],
\end{equation}
so that the total free energy is obtained as
\begin{equation}
{\beta \tilde{\cal F}_T(\sigma)\over V} =  {\Lambda^3\over 12 \pi^2}\left[ \ln\left(\varepsilon_w + \pi a^2 (\varepsilon_\perp -\varepsilon_w)nb\right) + \int_0^\infty {dt\over t}\left( {1\over (1+t)^{3\over 2}}
-\det[I+t(I+ \alpha \sigma)]^{-{1\over 2}}\right)\right] + {(n b)^2 k_BT \over 32~ K}
~{\rm Tr} ~\sigma^{-1}.
\end{equation}

The zero Matsubara frequency van der Waals free energy may now be expanded in terms of  $\alpha$,  the expansion up to order
$\alpha^2$ corresponding to  the  pairwise approximation for the van der Waals interaction.  Carrying out this expansion to 
third order in $\alpha$ yields up to terms independent of $\sigma$, 
\begin{eqnarray}
{\beta \tilde{\cal F}_T(\sigma)\over V} &=&  {\Lambda^3\over 12 \pi^2}\left[ {\alpha\over 3} {\rm Tr ~\sigma}
 -{\alpha^2\over 30} (2~{\rm Tr ~ \sigma^2} + ({\rm Tr~ \sigma})^2) + {\alpha^3\over 315}
 (8\  {\rm Tr}~\sigma^3  + 12\  {\rm Tr}~\sigma^2\ {\rm Tr}~\sigma
 + ({\rm Tr}~\sigma)^3) + {\cal O}(\alpha^4)\right ]  + \nonumber\\
 & & + ~{(n b)^2 k_BT \over 32~ K}
~{\rm Tr} ~\sigma^{-1} \label{fevdw}
\end{eqnarray}
It is interesting to compare our free energy functional at the order of $\alpha^2$ with that derived within the same mean-field approximation for  nematic interactions between polymer where  the 
pairwise interaction energy is given by \cite{Edwards}, 
\begin{equation}
{\cal E}_N = {\textstyle\frac12} u \int_0^L\!\!\int_0^L ds ds' ({\bf t}(s) \times {\bf t}(s'))^2 \delta({\bf x}(s) - {\bf x}(s')).
\end{equation}
The form of this interaction makes it energetically favorable for the polymers to align locally in a
parallel fashion. In the mean field approximation this gives a $\sigma$ dependent mean-field free energy of nematic interaction
\begin{equation}
{\beta \tilde{\cal F}_N(\sigma)\over V} = {1\over 2}\beta u [({\rm Tr} ~\sigma)^2 - {\rm Tr }~\sigma^2]
+{(n b)^2 k_BT \over 32~ K}
{\rm Tr} ~\sigma^{-1} ,\label{fenem}
\end{equation}
the part coming from the constrained polymer partition function being identical as a function of $\sigma$
in the two cases.  Recalling that as ${\rm Tr}~\sigma = nb$ is fixed we can identify an effective nematic interaction parameter for the van der Waals interaction by identifying the coefficients of ${\rm Tr}~\sigma^2$
in Eqs. (\ref{fevdw}) and ({\ref{fenem}) to obtain
\begin{equation}
u_e =   {k_B T \Lambda^3\alpha^2\over 90 \pi^2}.\label{eqedgu}
\end{equation}
The  effective nematic interaction is thus proportional to the square of the relative anisotropy polarizability $\alpha$  and thus crucially depends on the non-isotropic nature of the polymer material.  Note also that the sign of the interaction is always positive. This ties up with the fact that polarizable cylinders have a preference to align in a parallel fashion, thus favoring nematic order. However only stiff molecules with non-isotropic dielectric response, such as DNA or carbon nanotubes, can be expected to exhibit this particular tendency toward nematic ordering.

The variational  free energy density per unit volume $\tilde f(\sigma) = {{\tilde{\cal F}}_T(\sigma)/{V}},$ can be rewritten making a change of variables $\sigma\to nb ~\sigma'$ where now the constraint is ${\rm Tr} ~\sigma' = 1$
and the effective free energy per unit volume is obtained up to terms independent of $\sigma$ as
\begin{equation}
\beta \tilde f(\sigma')  = {\Lambda^3\over 12 \pi^2}\left[ \int_0^\infty {dt\over t}\left( {1\over (1+t)^{3\over 2}}
-\det[I+t(I+\gamma \sigma')]^{-{1\over 2}}\right) + c ~{\rm Tr} ~\sigma'^{-1}\right]. \label{fem}
\end{equation}
The dimensionless parameters $\gamma$ and $c$ are defined as
\begin{equation}
\gamma = nb\alpha = {\nu( \varepsilon_\parallel -\varepsilon_\perp)\over \varepsilon_w + \nu( \varepsilon_\parallel -\epsilon_w)} \qquad {\rm and} \qquad c ={3\over 8}{\pi^2 (n b)\over \Lambda^3 l_b b},
\label{gammanu}
\end{equation}
Here $\nu = n\pi a^2 b$ is the volume fraction of the polymer and $l_p=  K/ bk_B T$ is the dimensionless bare persistence length of the polymer in  units of the monomer length $b$.  From the definition it is clear that $0 < \gamma < ( \varepsilon_\parallel -\varepsilon_\perp)/ \varepsilon_\parallel $. This simple fact  is a crucial point arising from our analysis. The fact that the effective interaction parameter saturates at high volume concentrations means that a phase  transition cannot  be induced by simply increasing the concentration of polymer as in the case of models
with self-aligning-pairwise interactions \cite{Edwards}. Physically this result can be interpreted by the fact that n-body interactions tend to frustrate the ordering which is induced by the pairwise term.

In terms of the volume fraction we can write the variable $c$ as 
\begin{equation}
c = {3\pi\nu\over 8 \Lambda^3 a^2 b l_p}.
\end{equation}
Furthermore we assume that the monomer size $b$ and the radial size  of the polymer $a$ are of the same size. The cut-off $\Lambda$ in connection with vdW interactions is associated with the breakdown of the continuum dielectric description and is usually taken to be of the atomic size \cite{Ninham}. We assume the scaling form $\Lambda \simeq 2\pi \lambda/a$, where $\lambda$ is the ratio between the atomic and the macro-molecular sizes $\Lambda$ and $a$. This means that we can write
\begin{equation}
c \simeq {3 ~\nu a \over 64 \pi^2 \lambda^3~ b l_p} = {3 \over 64 \pi^2 \lambda^3}{\nu \over l_p} = {\cal C} ~\nu
\label{cnu}
\end{equation}
and we notice that the precise values of the parameters $a$ and $b$ can be absorbed into the redefinition of the persistence length  of the polymer, {\sl i.e.} $l_p \longrightarrow l_p (b/a)$. In the analysis that follows the dimensionless parameter $c$ is expected to be small as by definition $\nu<1$ and $\l_b >1$.   

\section{Nematic transition}

The parameter $\gamma$ is small in the limit where $\varepsilon_\parallel \approx \varepsilon_\perp$ and/or when the volume 
fraction $\nu$ is small. However for strong contrasts in $\varepsilon_\parallel$ and $\varepsilon_\perp$ it can be of order 1. It 
is convenient to measure the components of the polymer's dielectric tensor in terms of the dielectric constants of the solvent, we
write $ \varepsilon_\parallel = \varepsilon_w \epsilon_\parallel$ and $ \varepsilon_\parallel = \varepsilon_w \epsilon_\parallel$, 
which now gives
\begin{equation}
\gamma = \left({\epsilon_\parallel -\epsilon_\perp\over \epsilon_\parallel-1}\right)\left(1-{1\over 1 + \nu
(\epsilon_\parallel -1)}\right).
\end{equation}
Therefore $\gamma$  is a monotonic function of $\nu$ and the absolute value of $\gamma$ increases
upon increasing $\nu$.  In order to proceed we write the tensor  $\sigma$ as
\begin{equation}
\sigma = {1\over 3}  (I + Q) \label{sigq}
\end{equation}
where $Q$ is the  order parameter tensor 
\begin{equation}
Q= \begin{pmatrix}- S + T & 0 & 0 & \cr
                               0 & - S -T & 0 & \cr
                             0 & 0 & 2 S &
                             \end{pmatrix}.
                             \label{deford}
\end{equation}     
Here  $S$ is the standard uniaxial nematic scalar  order parameter and $T$ is the measure of the biaxality.

\begin{figure}[t]
\begin{center}
\vspace{0cm}
\includegraphics[angle=0,width=8.5cm]{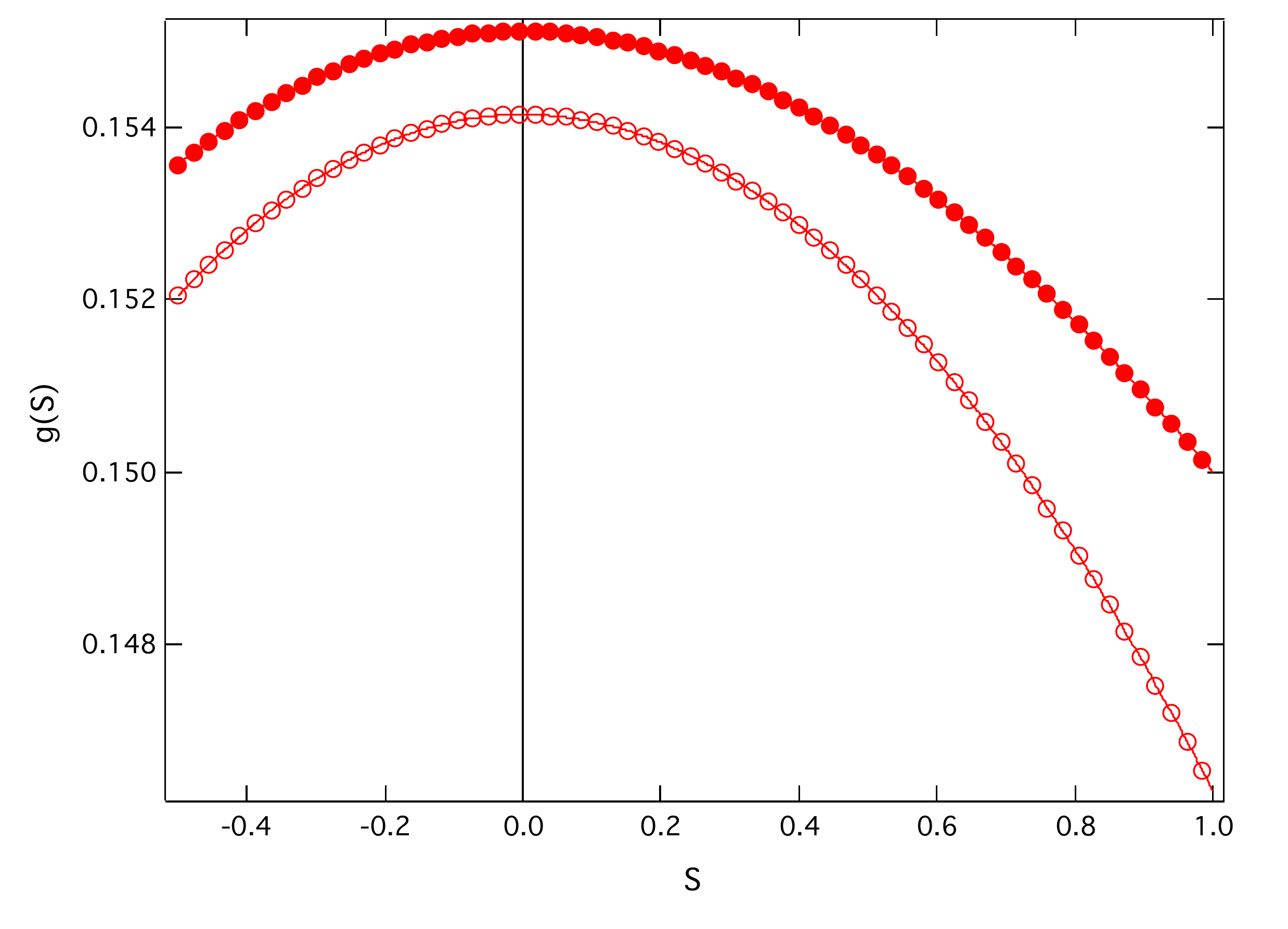}
\caption{The rescaled dependence of the zero frequency van der Waals energy on the order parameter, $g_{vdW}(S) = g(S, c= 0)$ (Eq. (\ref{g1})) for uniaxial ordering for the coupling parameter
$\gamma = 0.5$ compared with the expansions $g^{(3)}_{vdW}(S)$ of $g_{vdW}(S)$ to third order in $\gamma$ (Eq. (\ref{g3})).}
\label{g3vsg}
\end{center}
\end{figure}

At the pairwise level, the van der Waals interaction to order $\alpha^2$ can be mapped onto a nematic interaction Eq. (\ref{eqedgu}),  a uniaxial nematic alignment of the polymers is favoured and indeed Gupta and Edwards \cite{Edwards} found in their analysis that only these phases are thermodynamically stable. Thus in what follows we will examine the possibility of uniaxial ordering to all orders in $\alpha$ and thus set $T=0$ in the {\sl ansatz} Eq. (\ref{deford}). Note that in this case we must have $-{1/2}\leq S\leq 1$. While within this {\sl ansatz} positive $S$ corresponds to a normal nematic phase, the lower branch with negative $S$ implies excess material in the plane perpendicular to the nematic director and therefore represents a "discotic" phase. 

We first define the rescaled free energy Eq. (\ref{fem}) as
\begin{equation}
g(S) = {12 \pi^2 \beta \tilde f(\sigma')\over \Lambda^3} = \left[ \int_0^\infty {dt\over t}\left( {1\over (1+t)^{3\over 2}}
-\det[I+t(I+\gamma \sigma')]^{-{1\over 2}}\right) + c ~{\rm Tr} ~\sigma'^{-1}\right]. 
\end{equation}
Within the {\sl ansatz} Eq.  (\ref{deford}) it can be evaluated explicitly and is given by
\begin{equation}
g(S) = -2 + 2{\sqrt{3+\gamma(1-S)}\over \sqrt{3\gamma S}}\cos^{-1}\left({\sqrt{3+\gamma(1-S)}\over
\sqrt{3+ \gamma(1+2S)}}\right) + \ln\left( {1\over 3}(3 + \gamma(1+2S))\right) +{3 c\over 1+2S} + {6c\over 1-S}.\label{g1}
\end{equation}
The first three terms represent the zero Matsubara frequency contribution to the electromagnetic
field fluctuations. In case we would calculate the full Lifshitz expression in the case of magnetically inactive media, these 
three terms would need to be summed over all the Matsubara frequencies as in Eq. (\ref{fullmats}).

We denote the first three terms in Eq. (\ref{g1}) as $g_{vdW}(S)$, {\sl i.e.} $$g_{vdW}(S) = g(S, c =0).$$As one of the main aims of this paper is to investigate the effect of the full n-body van der 
Waals interactions it is interesting to compare this term with $g^{(2)}_{vdW}(S)$, which corresponds to the pairwise 
approximation, and can be derived from Eq. (\ref{g1}) or the expression Eq. (\ref{fevdw}) truncated at the second order in $\alpha$, which  gives
\begin{equation}
 g^{(2)}_{vdW}(S)= {\gamma\over 3} -{\gamma^2\over 90}( 5+4 S^2) + {\cal O}(\gamma^3).
 \end{equation}

\begin{figure}[t]
\begin{center}
\vspace{0cm}
\includegraphics[angle=0,width=8.5cm]{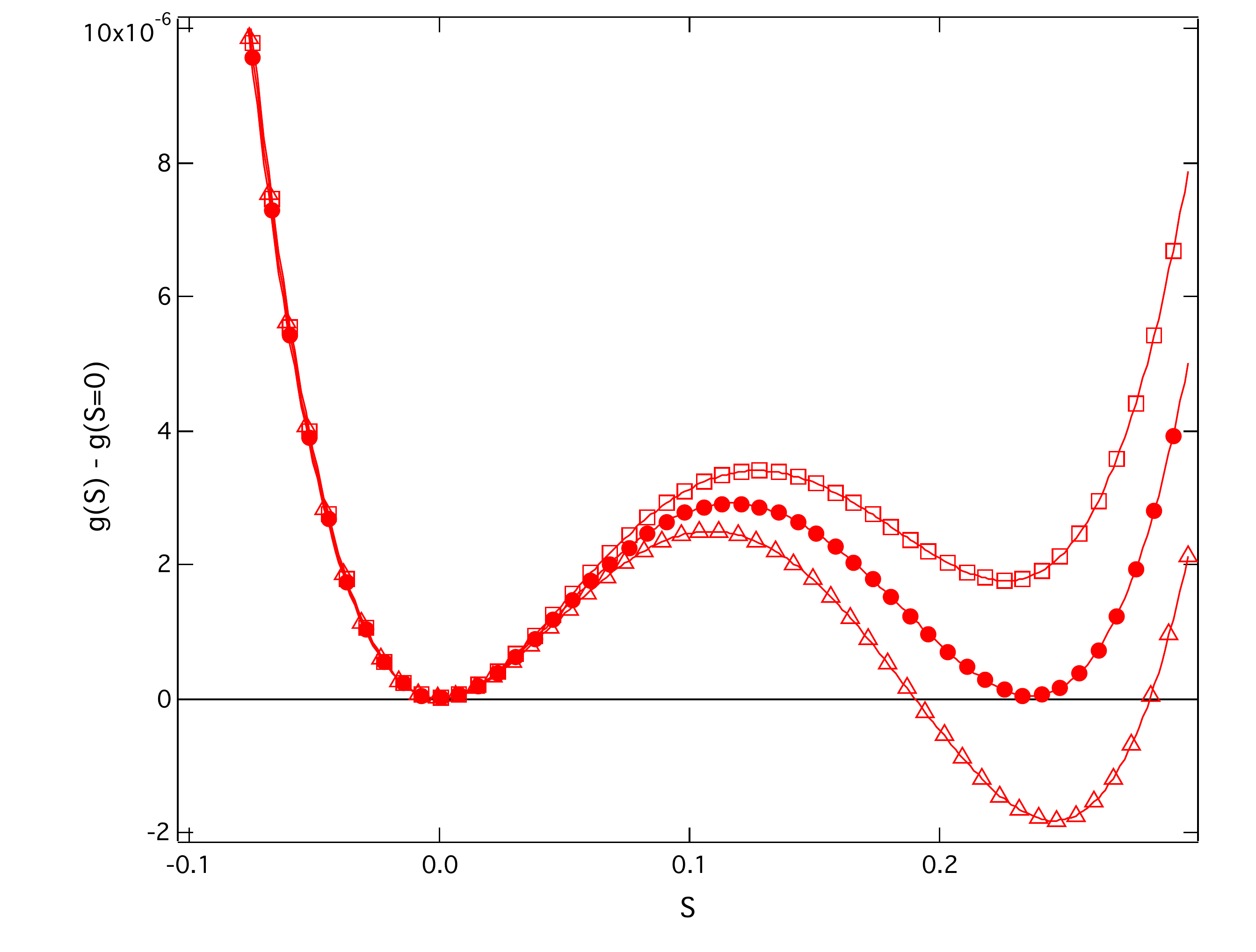}
\caption{The rescaled free energy difference (shifted for clarity by $g(S=0)$) $g(S)$ (Eq. (\ref{g1})) for uniaxial ordering for the coupling parameter $\gamma = 0.5$ for (from top to bottom) $c=0.000506$ (isotropic), $c=0.000504$ (coexistence between
isotropic and nematic phases) $c=0.000502$ (nematic phase).}
\label{transgam}
\end{center}
\end{figure}

The negative coefficient of $S^2$ clearly demonstrates that the van der Waals interactions  favor ordering 
at the pairwise level. However, one notes that at the pairwise level there is obviously no difference between the energy of nematic ordering (ordering along an axis) and discotic ordering (concentration in a plane) because the interaction is even in $S$. In the study of Edwards and Gupta \cite{Edwards} it was found that within the ordered phase it is the nematic phase  which has the lower free energy. The difference is due to the constraint part of $g(S)$, corresponding to the last two terms of Eq. (\ref{g1}), denoted by $g_{CO}(S)$, and given by
\begin{equation}
g_{CO}(S) = g(S) - g_{vdW}(S) = \frac{9 c~ (1+S)}{(1+2S) (1-S)}.
\end{equation}
This term favors nematic ordering over discotic ordering as can be seen by its Taylor expansion to fourth
order in $S$
\begin{equation}
g_{CO}(S)= 9c + 18c S^2 - 18c S^3 + 54c S^4 + {\cal O}(S^5).
\end{equation}
Here we can see how the symmetry of the van der Waals free energy between the nematic and discotic phases is broken by looking at the third order term in the interaction, corresponding to three body interactions. The total contribution to $g_{vdW}(S)$ to order $\gamma^3$ is then given by
 \begin{equation}
 g^{(3)}_{vdW}(S)= {\gamma\over 3} -{\gamma^2\over 90}( 5+4 S^2) +{\gamma^3\over 2835}(53 + 120 S^2 + 16 S^3) + {\cal O}(\gamma^4).\label{g3}
 \end{equation}
Therefore, for small $S$, where the term $S^2$ dominates, the three body interactions tend
to increase the van der Waals interaction energy and thus they reduce the tendency toward ordering.
Furthermore, we also see that the three body interaction favors discotic rather than nematic ordering. 

In Fig. (\ref{g3vsg}) we show the comparison of $g_{vdW}(S)$ with its approximation at third order in
$\gamma$, $g_{vdW}^{(3)}(S)$, for the value $\gamma=0.5$. We see that the approximation is 
very good at this value of $\gamma$, however the approximation tends to over estimate the
zero Matsubara frequency van der Waals energy at all values of $S$.  

As far as the phase diagram based on the assumption of nematic ordering is concerned, the rescaled free energy $g(S)$ for $\gamma=0.5$ is shown in Fig. (\ref{transgam}). We see that when the value of
$c$ is sufficiently lowered a first order nematic phase transition is induced. The transition occurs at 
$c^*\approx  0.000504$ and the order parameter $S$ jumps discontinuously from $0$ to $S^*\approx 0.24$. The complete dependence of the $c^*$ and $S^*$ on $\gamma$ is given on Fig. (\ref{cstarvgam}).

The way in which $c$ can be varied while keeping $\gamma$ constant is by changing the bare persistence length $l_p$ of the polymer.   The isotropic nematic transition occurs at a critical value
$l_p^*=  3\nu/64\pi^2 \lambda^3c^*$.  As $l_p^*$ is further increased, the order parameter $S$  increases and eventually tends to $1$.

\begin{figure}
\begin{center}

\includegraphics[angle=0,width=8.5cm]{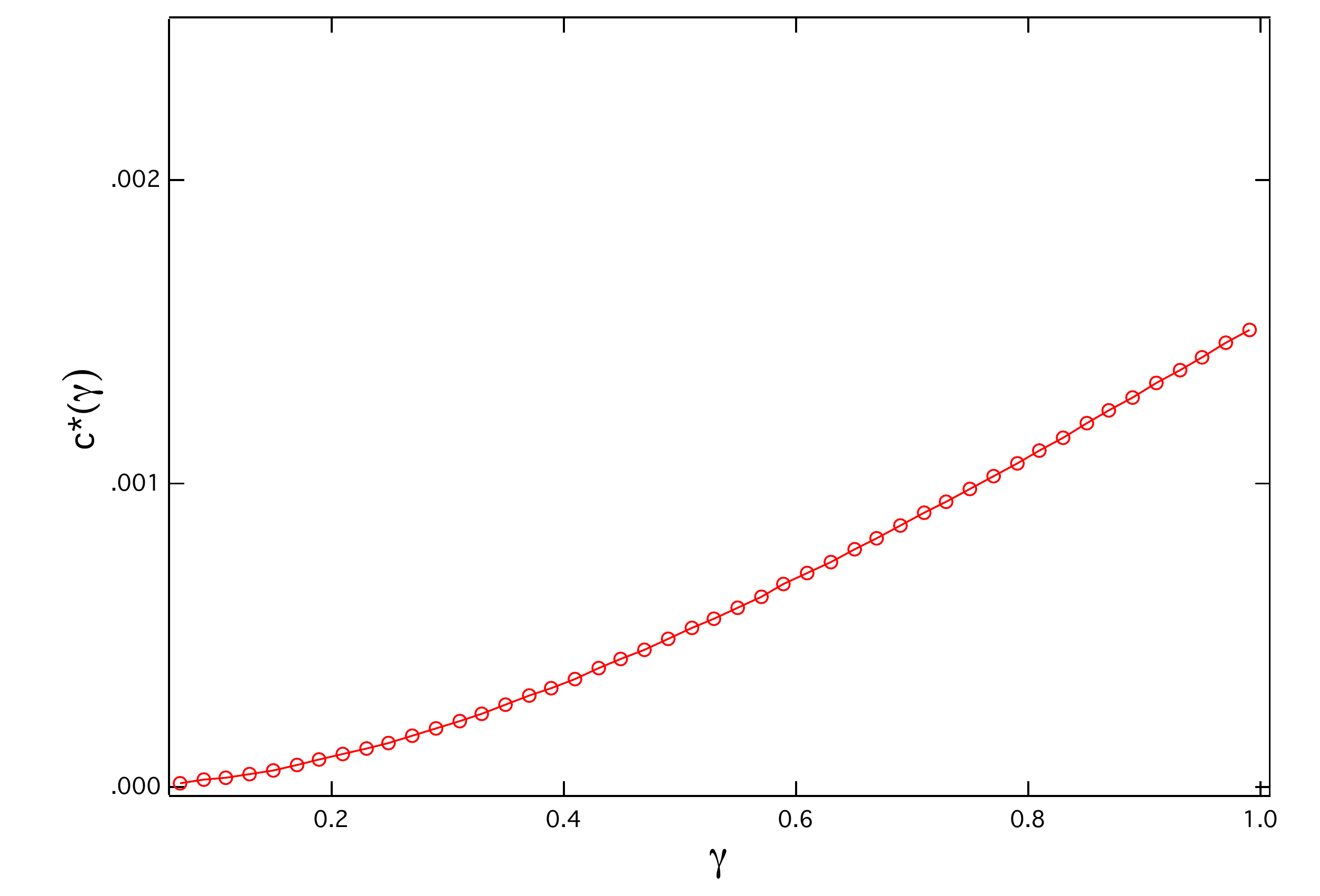}
\includegraphics[angle=0,width=8.5cm]{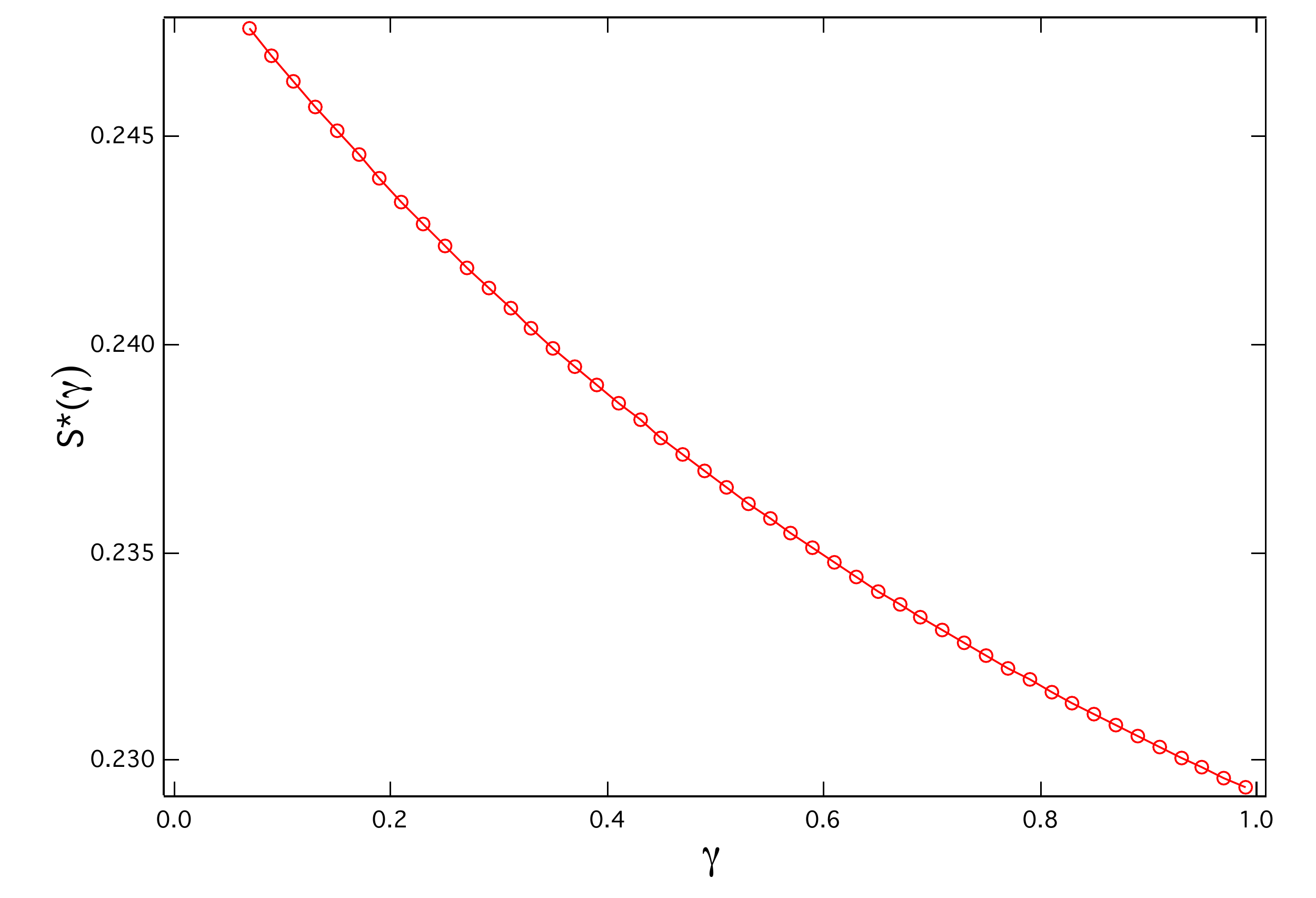}

\caption{The dependence of the $c^*$ and $S^*$ at the nematic transition on the interaction parameter $\gamma = {\nu( \varepsilon_\parallel -\varepsilon_\perp)\over \varepsilon_w + \nu( \varepsilon_\parallel -\epsilon_w)} $.}
\label{cstarvgam}
\end{center}
\end{figure}

Inspite of the insights gained by this partitioning of the volume fraction dependence of $\gamma$ and $c$, it
is necessary to investigate also the complete dependence of the free energy on the polymer density or equivalently its volume fraction.
This dependence is hidden in $\gamma = \gamma(\nu)$, Eq. (\ref{gammanu}), and $c = c(\nu)$, Eq. (\ref{cnu}).

\begin{figure}
\begin{center}

\includegraphics[angle=0,width=8.5cm]{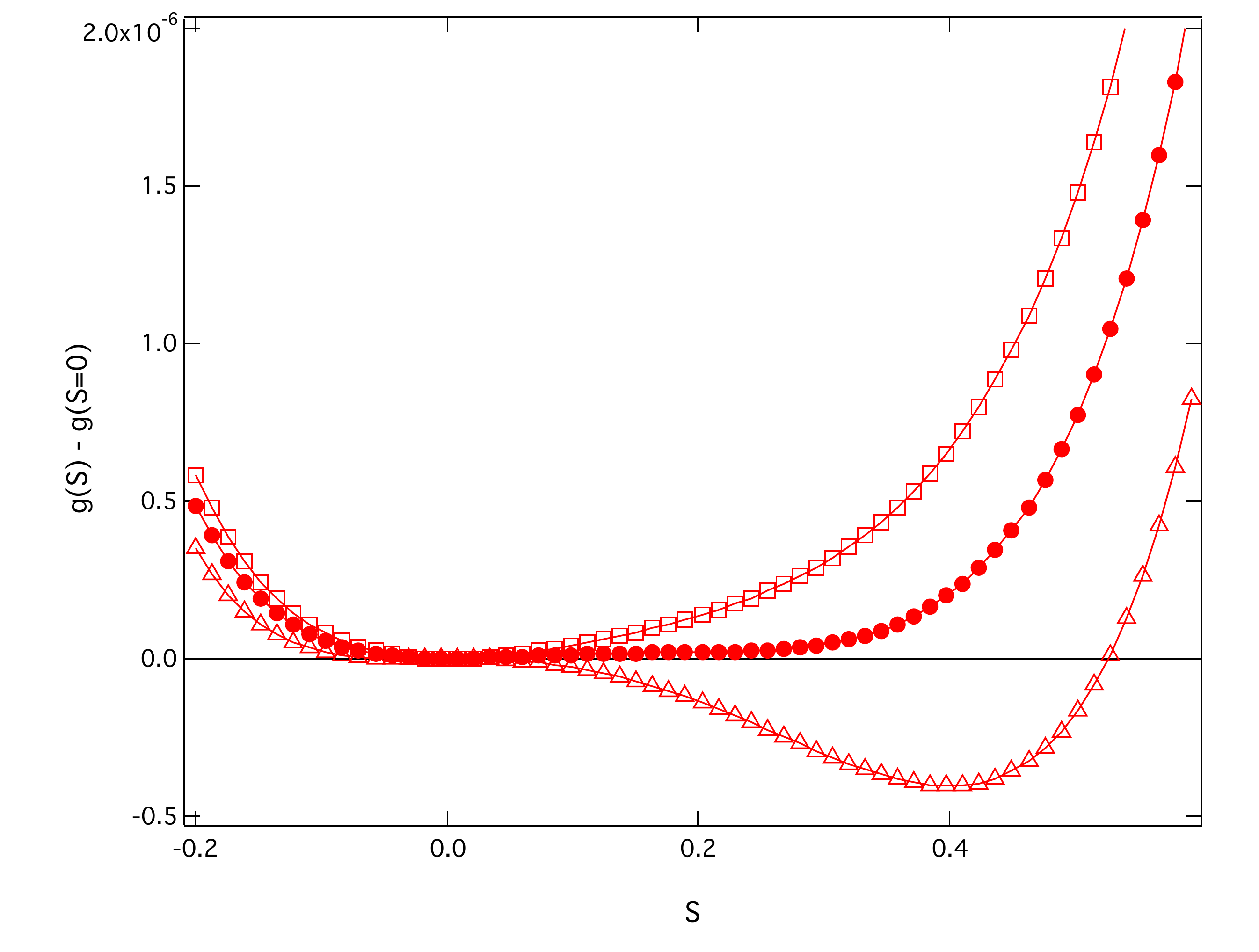}

\caption{The dependence of the free energy difference $g(S) - g(S=0)$ on the order parameter $S$ for various values of the volume fraction. Top
to bottom $\nu = 0.63, 0.655, 0.68$. We have taken $\varepsilon_\parallel = 4$, $\varepsilon_\perp = 3$ and $\varepsilon_w = 80$ and ${\cal C} = 2 \times 10^{-6}$. The value for $\cal C$ corresponds to $l_b = 2.5$ and $\lambda = 10$. }
\label{freen}
\end{center}
\end{figure}

Fig. (\ref{freen}) shows the order parameter dependence of the total free energy Eq. (\ref{g1}) for three different values of the volume fraction $\nu$ contained in $\gamma = \gamma(\nu)$ and $c = c(\nu)$. The dielectric constants were taken as $\varepsilon_\parallel = 4$, $\varepsilon_\perp = 3$ and $\varepsilon_w = 80$, which corresponds to a polymer with a hydrophobic core in an aqueous solution. The persistence length was taken as 2.5 in the units of the monomer size and the radius of the polymer was taken as ten times the atomic size cutoff,  $\lambda = 10$. 

For the chosen set of parameters the nematic transition occurs at $\nu = 0.655$, where the minimum of the free energy is displaced from its disordered value $S = 0$ to an ordered nematic state with $S \neq 0$. The critical concentration obviously depends on the exact values of the persistence length, the radius of the polymer and its dielectric anisotropy.

\section{Discussion}

We explored the consequences of the orientationally dependent anisotropic vdW interactions on the ordering transition
of semi flexible polymers. A theory {\sl exact} to all orders in the non-pairwise additive vdW interactions was 
derived and solved on the mean-field constant density level. We show that indeed there exists an orientational ordering transition engendering a nematic polymer phase. 

Through we limited ourselves to the zero Matsubara frequency term, the approach can be straightforwardly generalized
to include the full anisotropic Lifshitz expression \cite{vdWbooks}.  In this case the free energy Eq. (\ref{g1}) needs to be rewritten as a full Matsubara frequency sum of the form
\begin{equation}
g(S, \nu) = {\sum_{n = 0}^{\infty}}' \left( -2 + 2{\sqrt{3+\gamma_n (1-S)}\over \sqrt{3\gamma_n  S}}\cos^{-1}\left({\sqrt{3+\gamma_n (1-S)}\over
\sqrt{3+ \gamma_n (1+2S)}}\right) + \ln\left( {1\over 3}(3 + \gamma_n (1+2S))\right)\right)  + \frac{9 c ~(1 + S)}{(1+2S)(1-S)},\label{g1-1}
\end{equation}
where now
\begin{equation}
\gamma_n = \gamma(\nu, \imath \xi_n) = {\nu~ ( \varepsilon_\parallel(\imath \xi_n) -\varepsilon_\perp(\imath \xi_n))\over \varepsilon_w(\imath \xi_n) + \nu~( \varepsilon_\parallel(\imath \xi_n) -\epsilon_w(\imath \xi_n))} \qquad {\rm and} \qquad c = {3\pi\nu\over 8 ~\Lambda^3 a^2 b l_p}  = {3 \over 64 \pi^2 \lambda^3}{\nu \over l_p}.
\end{equation}
Depending on the dielectric anisotropy $\varepsilon_\parallel(\imath \xi_n) -\varepsilon_\perp(\imath \xi_n)$ at every Matsubara frequency, each term in the sum makes a contribution to the total ordering free energy. This means that the vdW part of the free energy in general increases and therefore displaces the ordering transition towards smaller densities. One should note that the overall density dependence of Eq. (\ref{g1-1}) becomes quite complicated in this case as partly (the first three terms) it depends on the frequency dependent dielectric response, and partly it doesn't (the last term). We also 
notice that the temperature dependence of the free energy upon including the non-zero Matsubara frequencies becomes
much more complex. 

The total free energy thus contains two separate contributions: the Matsubara sum corresponding to the anisotropic vdW interactions and the last term stemming from the "constraint" free energy, that connects the dielectric anisotropy of the polymer chain with its configuration in space. It is the last term that can not be evaluated precisely because it contains a spatial cutoff $\Lambda$ due to the continuum nature of the electrostatic field description. While this is not a major impediment regarding the scaling properties and the existence of the transition, it can introduce afterthoughts regarding the exact numerical values for the transition density. Based on numerical calculations we conclude that the polymer does not have to be very stiff in order to show the vdW driven orientational ordering transition. 


The mean-field approach outlined here for the case of orientationally dependent van der Waals interactions can of course be easily extended and applied to other cases. Mean-field electrostatic interactions would be a possible candidate.

\section{Acknowledgement}

D.S.D. acknowledges support from the Institut Universitaire de France.  R.P. acknowledges support from ARRS through the program P1-0055 and the research project J1-0908 as well as the University of Toulouse for a one month position of {\em Professeur invit\' e}. 


\end{document}